\documentclass[italian,english]{article}
\usepackage[latin9]{inputenc}
\usepackage{color}
\usepackage{float}
\usepackage{amstext}
\usepackage{graphicx}

\makeatletter
\newcommand{\lyxaddress}[1]{
\par {\raggedright #1
\vspace{1.4em}
\noindent\par}
}

\makeatother

\usepackage{babel}

\begin{document}

\title{\textbf{Tuning the stochastic background of gravitational waves using
the WMAP data}}

\author{\textbf{Christian Corda}}

\maketitle

\lyxaddress{\begin{center}
International Institute for Theoretical Physics and Advanced Mathematics
Einstein-Galilei, Via Santa Gonda, 14 - 59100 PRATO, Italy
\par\end{center}}

\lyxaddress{\begin{center}
\textit{E-mail addresses:} \textcolor{blue}{cordac.galilei@gmail.com}
\par\end{center}}
\begin{abstract}
The cosmological bound of the stochastic background of gravitational
waves is analyzed with the auxilium of the WMAP data, differently
from lots of works in literature, where the old COBE data were used.
From our analysis, it will result that the WMAP bounds on the energy
spectrum and on the characteristic amplitude of the stochastic background
of gravitational waves are greater than the COBE ones, but they are
also far below frequencies of the earth-based antennas band. At the
end of this letter a lower bound for the integration time of a potential
detection with advanced LIGO is released and compared with the previous
one arising from the old COBE data. Even if the new lower bound is
minor than the previous one, it results very long, thus for a possible
detection we hope in the LISA interferometer and in a further growth
in the sensitivity of advanced projects.
\end{abstract}
The design and construction of a number of sensitive detectors for
gravitational waves (GWs) is underway today. There are some interferometers
like the VIRGO detector, being built in Cascina near Pisa, Italy,
by a joint Italian-French collaboration, the GEO 600 detector being
built in Hanover, Germany by a joint Anglo-Germany collaboration,
the two LIGO detectors being built in the United States (one in Hanford,
Washington and the other in Livingston, Louisiana) by a joint Caltech-Mit
collaboration, and the TAMA 300 detector, being built near Tokyo,
Japan. There are many bar detectors currently in operation too, and
several interferometers and bars are in a phase of planning and proposal
stages (for the current status of GWs experiments see \cite{key-1,key-2}).

The results of these detectors will have a fundamental impact on astrophysics
and gravitation physics. There will be several experimental data to
be analyzed, and theorists will be forced to interact with lots of
experiments and data analysts to extract the physics from the data
stream.

In this letter the stochastic background of GWs (SBGWs) \cite{key-3,key-4,key-5},
which is, in principle, a possible target of these experiments, is
analyzed with the auxilium of the WMAP data \cite{key-6,key-7}. We
emphasize that, in general, in previous works in literature about
the SBGWs, old COBE data were used (see \cite{key-3,key-4,key-5,key-8,key-10}
for example).

From our analysis, it will result that the WMAP bounds on the energy
spectrum and on the characteristic amplitude of the SBGWs are greater
than the COBE ones, but they are also far below frequencies of the
earth-based antennas band. At the end of this letter a lower bound
for the integration time of a potential detection with advanced LIGO
is released and compared with the previous one arising from the old
COBE data. Even if the new lower bound is minor than the previous
one, it results very long, thus for a possible detection we hope in
the LISA interferometer and in a further growth in the sensitivity
of advanced projects.

The strongest constraint on the spectrum of the relic SBGWs in the
frequency range of ground based antennas like bars and interferometers,
which is the range $10Hz\leq f\leq10^{4}Hz$, comes from the high
isotropy observed in the Cosmic Background Radiation (CBR).

The fluctuation $\Delta T$ of the temperature of CBR from its mean
value $T_{0}=2.728$ K varies from point to point in the sky \cite{key-6,key-7},
and, since the sky can be considered the surface of a sphere, the
fitting of $\Delta T$ is performed in terms of a Laplace series of
spherical harmonics

\begin{equation}
\frac{\Delta T}{T_{0}}(\hat{\Omega})=\sum_{l=1}^{\infty}\sum_{m=-l}^{l}a_{lm}Y_{lm}(\hat{\Omega}),\label{eq: fluttuazioni CBR}\end{equation}

and the fluctations are assumed statistically independent ($<a_{lm}>=0$,
$<a_{lm}a_{l'm'}^{*}>=C_{l}\delta_{ll'}\delta_{mm'}$). In eq. (\ref{eq: fluttuazioni CBR})
$\hat{\Omega}$ denotes a point on the 2-sphere while the $a_{lm}$
are the multipole moments of the CBR. For details about the definition
of statistically independent fluctations in the context of the temperature
fluctations of CBR see \cite{key-11}.

The WMAP data \cite{key-6,key-7} permit a more precise determination
of the rms quadrupole moment of the fluctations than the COBE data

\begin{equation}
Q_{rms}\equiv T(\sum_{m=-2}^{2}\frac{|a_{2m}|^{2}}{4\pi})^{\frac{1}{2}}=8\pm2\mu K,\label{eq: Q rms}\end{equation}

while in the COBE data we had \cite{key-3,key-14}

\begin{equation}
Q_{rms}=14.3_{-3.3}^{+5.2}\mu K.\label{eq: Q rms COBE}\end{equation}
 A connection between the fluctuation of the temperature of the CBR
and the SBGWs derives from the \textit{Sachs-Wolfe effect} \cite{key-8,key-9}.
Sachs and Wolfe showed that variations in the density of cosmological
fluid and GWs perturbations result in the fluctuation of the temperature
of the CBR, even if the surface of last scattering had a perfectly
uniform temperature \cite{key-9}. In particular the fluctuation of
the temperature (at the lowest order) in a particular point of the
space is

\begin{equation}
\frac{\Delta T}{T_{0}}(\hat{\Omega})=\frac{1}{2}\int_{nullgeodesic}d\lambda\frac{\partial}{\partial\eta}h_{rr}.\label{eq: null geodesic}\end{equation}

The integral in eq. (\ref{eq: null geodesic}) is taken over a path
of null geodesic which leaves the current spacetime point heading
off in the direction defined by the unit vector $\hat{\Omega}$ and
going back to the surface of last scattering of the CBR.

Here $\lambda$ is a particular choice of the affine parameter along
the null geodesic. By using conformal coordinates, we have for the
metric perturbation

\begin{equation}
\delta g_{ab}=R^{2}(\eta)h_{ab},\label{eq: metric perturbation}\end{equation}

and $r$ in eq. (\ref{eq: null geodesic}) is a radial spatial coordinate
which goes outwards from the current spacetime point. The effect of
a long wavelenght GW is to shift the temperature distribution of CBR
from perfect isotropy. Because the fluctations are very small ($<\Delta T/T_{0}>\leq5*10^{-5}$
\cite{key-6,key-7}), the perturbations caused by the relic SBGWs
cannot be too large. 

The WMAP results give rather tigh constraints on the spectrum of the
SBGWs at very long wavelenghts. In \cite{key-3} we find a constraint
on $\Omega_{gw}(f)$ which derives from the COBE observational limits,
given by

\begin{equation}
\Omega_{gw}(f)h_{100}^{2}<7*10^{-11}(\frac{H_{0}}{f})^{2}\textrm{ for }H_{0}<f<30H_{0}.\label{eq: limite COBE}\end{equation}

Now the same constraint will be obtained from the WMAP data. Because
of its specific polarization properties, relic SBGWs should generate
particular polarization pattern of CBR anisotropies, but the detection
of CBR polarizations is not fulfiled today \cite{key-13}. Thus an
indirect method will be used. We know that relic GWs have very long
wavelenghts of Hubble radius size, so the CBR power spectrum from
them should manifest fast decrease at smaller scales (hight multipole
moments). But we also know that scalar modes produce a rich CBR power
spectrum at large multipole moments (series of acoustic peaks, ref.
\cite{key-6,key-7}). Then the properties of tensor modes of cosmological
perturbations of spacetime metric can be extract from observational
data using angular CBR power spectrum combined with large scale structure
of the Universe. One can see (fig. 1 ) that in the range $2\leq l\leq30$
(the same used in \cite{key-3}, but with the old COBE data \cite{key-14})
scalar and tensor contributions are comparable. From \cite{key-6,key-7}
the WMAP data give for the tensor/scala ratio $r$ the constraint
$r<0.9$. \foreignlanguage{italian}{(}$r<0.5$ \foreignlanguage{italian}{in
the COBE data, ref. \cite{key-14}); Novosyadly and Apunevych} obtained
$\Omega_{scalar}(H_{0})<2.7*10^{-9}$ \cite{key-13}. Thus, if one
remembers that, at order of Hubble radius, the tensorial spectral
index is $-4\leq n_{t}\leq-2$ , it results

\begin{equation}
\Omega_{gw}(f)h_{100}^{2}<1.6*10^{-9}(\frac{H_{0}}{f})^{2}\textrm{ for }H_{0}<f<30H_{0},\label{eq: limite WMAP}\end{equation}

which is greater than the COBE data result of eq. (\ref{eq: limite COBE}).

We emphasize that the limit of eq. (\ref{eq: limite WMAP}) is not
a limit for any GWs, but only for relic ones of cosmological origin,
which were present at the era of the CBR decoupling. Also, the same
limit only applies over very long wavelenghts (i.e. very low frequencies)
and it is far below frequencies of the Virgo - LIGO band.

\begin{figure}
\includegraphics[scale=0.9]{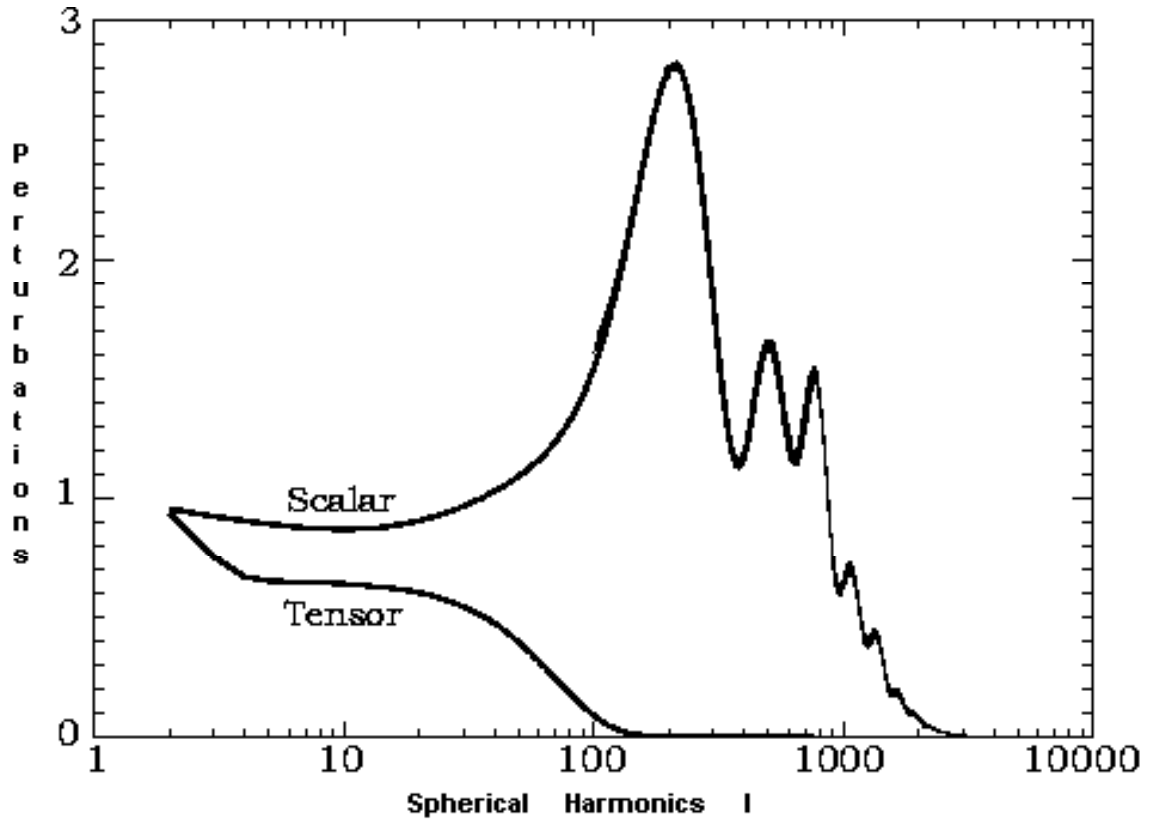}

\caption{The tensor to scalar ratio}

\end{figure}

The primordial production of the relic SBGWs has been analyzed in
\cite{key-3,key-4,key-5,key-15,key-16}, where it has been shown that
in the range $10^{-15}Hz\leq f\leq10^{10}Hz$ the spectrum is flat
and proportional to the ratio

\begin{equation}
\frac{\rho_{ds}}{\rho_{Planck}}\approx10^{-12}.\label{eq: rapporto densita' primordiali}\end{equation}

WMAP observations put strongly severe restrictions on the spectrum,
as we discussed above. In fig. 2 the spectrum $\Omega_{gw}$ is mapped:
the amplitude (determined by the ratio $\frac{\rho_{ds}}{\rho_{Planck}}$)
has been chosen to be \textit{as large as possible, consistent with
the WMAP constraint} (\ref{eq: limite WMAP}). Nevertheless, because
the spectrum falls off $\propto f^{-2}$ at low frequencies \cite{key-3,key-4,key-5},
this means that today, at Virgo and LISA frequencies, indicated in
fig. 2,

\begin{center}
\begin{equation}
\Omega_{gw}(f)h_{100}^{2}<9*10^{-13},\label{eq: limite spettro WMAP}\end{equation}

\par\end{center}

while using the COBE data it was

\begin{center}
$\Omega_{gw}(f)h_{100}^{2}<8*10^{-14}$(refs. \cite{key-3,key-14}).
\par\end{center}

It is intersting to calculate the correspondent strain at $\approx100Hz$,
where interferometers like Virgo and LIGO have a maximum in sensitivity.
The well known equation for the characteristic amplitude \cite{key-3,key-8}
can be used:

\begin{equation}
h_{c}(f)\simeq1.26*10^{-18}(\frac{1Hz}{f})\sqrt{h_{100}^{2}\Omega_{gw}(f)},\label{eq: legame ampiezza-spettro}\end{equation}
obtening

\begin{equation}
h_{c}(100Hz)<1.7*10^{-26}.\label{eq: limite per lo strain}\end{equation}

Then, because for ground-based interferometers a sensitivity of the
order of $10^{-22}$ is expected at $\approx100Hz$, four order of
magnitude have to be gained in the signal to noise ratio \cite{key-17,key-18}.
Let us analyze smaller frequencies too. The sensitivity of the Virgo
interferometer is of the order of $10^{-21}$ at $\approx10Hz$ \cite{key-17}
and in that case it is 

\begin{equation}
h_{c}(100Hz)<1.7*10^{-25}.\label{eq: limite per lo strain2}\end{equation}

For a better understanding of the difficulties on the detection of
the SBGWs a lower bound for the integration time of a potential detection
with advanced LIGO is released. For a cross-correlation between two
interferometers the signal to noise ratio (SNR) increases as \cite{key-3,key-4,key-5,key-19}

\begin{equation}
(SNR)=\sqrt{2T}\frac{H_{0}^{2}}{5\pi^{2}}\sqrt{\int_{0}^{\infty}df\frac{\Omega_{gw}^{2}(f)\gamma^{2}(f)}{f^{6}P_{1}(|f|)P_{2}(|f|)}}.\label{eq: SNR2}\end{equation}

where $P_{i}(|f|)$ is the one-sided power spectral density of the
$i$ detector \cite{key-19} and $\gamma(f)$ the well known overlap-reduction
function \cite{key-19,key-20}. Assuming two coincident coaligned
detectors $(\gamma(f)=1)$ with a noise of order $10^{-48}/Hz$ (i.e.
a typical value for the advanced LIGO sensitivity \cite{key-21})
one gets $(SNR)\sim1$ for $\sim3*10^{5}years$ using our result $\Omega_{gw}(f)h_{100}^{2}\sim9*10^{-13}$
while it is $(SNR)\sim1$ for $\sim3*10^{7}years$ using previous
COBE result $\Omega_{gw}(f)h_{100}^{2}\sim8*10^{-14}$. Since the
overlap reduction function degrades the SNR, these results can be
considered a solid upper limit for the advanced LIGO configuration
for the two different values of the spectrum. 

The sensitivity of the LISA interferometer will be of the order of
$10^{-22}$ at $10^{-3}\approx Hz$ \cite{key-22} and in that case
it is 

\begin{equation}
h_{c}(100Hz)<1.7*10^{-21}.\label{eq: limite per lo strain3}\end{equation}

Then a stochastic background of relic gravitational waves could be
in principle detected by the LISA interferometer. We also hope in
a further growth in the sensitivity of advanced projects.

We emphasize that the assumption that all the tensorial perturbation
in the Universe are due to a SBGWs is quit strong, but our results
(\ref{eq: limite spettro WMAP}), (\ref{eq: limite per lo strain}),
(\ref{eq: limite per lo strain2}) and (\ref{eq: limite per lo strain3})
can be considered like upper bounds. 

Reasuming in this letter the SBGWs has been analyzed with the auxilium
of the WMAP data, while previous works in literature, used the old
COBE data, seeing that the predicted signal for these relic GWs is
very weak. From our analysis it resulted that the WMAP bound on the
energy spectrum and on the characteristic amplitude of the SBGWs are
greater than the COBE ones, but they are also far below frequencies
of the earth-based antennas band. In fact the integration time of
a potential detection with advanced interferometers is very long,
thus, for a possible detection we have to hope in a further growth
in the sensitivity of advanced ground based projects and in the LISA
interferometer.

\begin{figure}[H]
\includegraphics[scale=0.9]{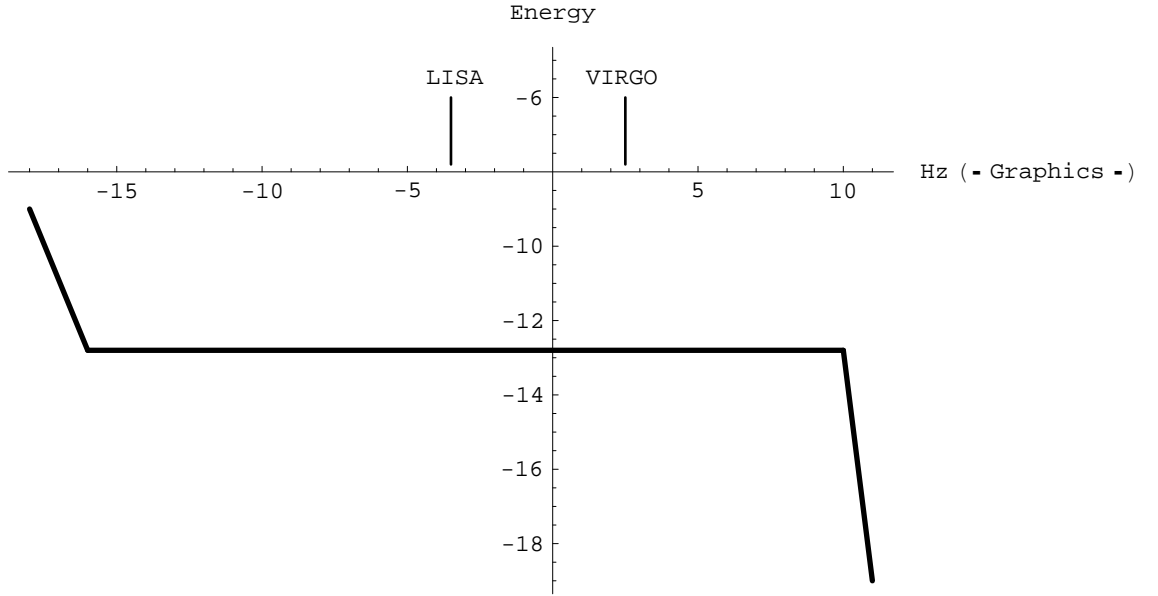}

\caption{The spectrum of relic gravitons in inflationary models is flat over
a wide range of frequencies. The horizontal axis is $\log_{10}$ of
frequency, in Hz. The vertical axis is $\log_{10}\Omega_{gr}$. The
inflationary spectrum rises quickly at low frequencies (wave which
rentered in the Hubble sphere after the Universe became matter dominated)
and falls off above the (appropriately redshifted) frequency scale
$f_{max}$ associated with the fastest characteristic time of the
phase transition at the end of inflation. The amplitude of the flat
region depends only on the energy density during the inflationary
stage; we have chosen the largest amplitude consistent with the WMAP
constrain discussed earlier: $\Omega_{gr}(f)h_{100}^{2}<1.6*10^{-9}$
at $10^{-18}Hz$. This means that at Virgo and Lisa frequencies, $\Omega_{gr}(f)h_{100}^{2}<9*10^{-13}$}

\end{figure}

\section*{Acknowledgements}

I would like to thank Maria Felicia De Laurentis and Giancarlo Cella
for helpful advices during my work. I thank the referee for its interest
in my work and for precious advices and comments that allowed to improve
this paper. The European Gravitational Observatory (EGO) consortium
has also to be thanked for the using of computing facilities.


\begin{thebibliography}{22}
\bibitem{key-1}http://www.ligo.org/pdf\_public/camp.pdf

\bibitem{key-2}http://www.ligo.org/pdf\_public/hough02.pdf

\bibitem{key-3}B. Allen - Proceedings of the Les Houches School on
Astrophysical Sources of Gravitational Waves, eds. Jean-Alain Marck
and Jean-Pierre Lasota (Cambridge University Press, Cambridge, England
1998).

\bibitem{key-4}\foreignlanguage{italian}{L. P. Grishchuk and others
- Phys. Usp. 44 1-51 (2001)}

\bibitem{key-5}\foreignlanguage{italian}{L. P. Grishchuk and others
- Usp. Fiz. Nauk 171 3 (2001)}

\bibitem{key-6}\foreignlanguage{italian}{C. L. Bennet and others
- ApJS \textbf{148} 1}

\bibitem{key-7}\foreignlanguage{italian}{D. N. Spergel and others
- ApJS \textbf{148} 195}

\bibitem{key-8}M. Maggiore - Physics Reports \textbf{331}, 283-367
(2000)

\bibitem{key-9}R. K. Sachs and A. M. Wolfe - \foreignlanguage{italian}{ApJ
147, 73 (1967)}

\begin{flushleft}
\bibitem{key-10}\foreignlanguage{italian}{D. Babusci and others  -
Virgo DAD - www.virgo.infn.it/documents/DAD/stochastic background}
\par\end{flushleft}

\bibitem{key-11}\foreignlanguage{italian}{S. N. Shore - {}``The
tapestry of modern astrophysics'' - Wiley Interscience (2003)}

\bibitem{key-12}\foreignlanguage{italian}{C. W. Misner, K. S. Thorne
and J. A. Wheeler - {}``Gravitation'' - W.H.Feeman and Company -
1973}

\bibitem{key-13}\foreignlanguage{italian}{B. Novosyadlyj and S. Apunevych
- proceedings of international confernce {}``Astronomy in Ukraine
- Past, Present, Future'' - Main Astronomical Observatory (2004)}

\bibitem{key-14}J. P. Zibin, D. Scott and M. White \foreignlanguage{italian}{-
arXiv:astro-ph/9904228}

\bibitem{key-15}B. Allen and A.C. Ottewill - Phys. Rev. D \textbf{56}
545-563 (1997)

\bibitem{key-16}B. Allen - Phys. Rev. D \foreignlanguage{italian}{\textbf{3}}\textbf{7},
2078 (1988)

\bibitem{key-17}F. Acernese et al. (the Virgo Collaboration) - Class.
Quant. Grav. \textbf{23} 19 S635-S642 (2006)

\bibitem{key-18}S. Hild (for the LIGO Scientific Collaboration) -
Class. Quant. Grav. \textbf{23} 19 S643-S651 (2006)

\bibitem{key-19}B. Allen and J. P. Romano - Phys. Rev. D \foreignlanguage{italian}{\textbf{59}
102001 (1999)}

\bibitem{key-20}E. E. Flanagan - Phys. Rev. D \foreignlanguage{italian}{\textbf{48}
2389 (1993)}

\bibitem{key-21}K. G. Arun, B. R. Iver, B. S. Sathyaprakash, and
P. A. Sundararajan - Phys. Rev. D \foreignlanguage{italian}{\textbf{71}
084008 (2005)}

\bibitem{key-22}www.lisa.nasa.org; www.lisa-scienze.org 
\end{thebibliography}
\end{document}